# Morphology of Flow Patterns Generated by Viscous Fingering from Miscible Fluids

Lang Xia

Email: langxia.org@gmail.com

**Abstract:** A rest fluid displaced by a less viscous fluid in a porous medium triggers the so-called Saffman-Taylor instability at their contact front and hence forms complicated finger-like patterns. When the two fluids are miscible, the surface tension at their contact front vanishes, leaving the variation in viscosity dominant the contribution to the instability. The phenomena, named viscous fingering, can be studied by the analogy of a single-phase flow in the Hele-Shaw cell, a quasi-two-dimensional rectilinear plane. Viscous fingering produces complex geometrical patterns. They are important not only in industry but also in mathematics. Theoretical analyses on the evolution of the finger-like patterns in miscible fluids are still expanding (it is referred to the morphology of the flow patterns). Here we study the morphology of the finger-like patterns within some simple concepts in differential geometry, in which surfaces and curves are analog to the profiles of the concentration and its contours, respectively. We thus can investigate the fingering phenomena through the geometrical perspective in which various results in differential geometry are immediately applicable.

## 1. Introduction

Mixing flows of two or more fluids producing rich phase diagrams occur ubiquitously in the natural world. However, the phenomena accompanying with mixing flows had long been taken for granted. The developing of techniques to reproduce the mixing flows in the laboratory had seldom been recorded until the dawn of the twentieth century. The earliest mixing techniques could be dated back to the sixteenth century, which the manual mixing was being supplemented by mechanical agitations (Marshall and Bakker 2004). Like many other natural phenomena, the fluid mixing can only be understood by performing proper scientific experiments and employing related mathematical tools. Topics involved in the mixing flows gained much interest in the second half of the twentieth century. Laboratory experiments were able to replicate the mixing phenomena found in the natural world (Read 1984, Diment and Watson 1985, Nittmann, Daccord et al. 1985, Kusch and Ottino 1992). On the other hand, the mixing flows were also simulated numerically with the developing of modern algorithms and computers (Ashurst 1977, Nittmann, Daccord et al. 1985,

Bensimon, Kadanoff et al. 1986, Tan and Homsy 1988, Brennen 2005, Dong, Yan et al. 2011).

The mixing phenomena featuring the formation of finger-like or dendritic patterns usually take place at the front of two or more fluids of different physical properties, such as density, viscosity, or temperature. Mixing was defined as the reduction of inhomogeneity of the fluids in order to achieve the desired process result (Marshall and Bakker 2004). When one of the two different fluids is driven by another, the advection carries fluid particles (or physical properties) along the streamline while the diffusion spreads them across the streamline. The interplaying of the advection and diffusion effects triggers mixing process and possibly forms intriguing patterns. Therefore, mixing may be enhanced by increasing the variations of the density, velocity, viscosity and so forth. Investigations on the mixing flows have been spanning into many different fields. Besides the work in classical fluid mechanics (Stone, Stroock et al. 2004), biogenic mixing was studied in a large-scale in the ocean biochemistry (Katija and Dabiri 2009). It helps to understand the nutrient transport in ocean globally; also, the theory of fluid mixing was applied to the study of galaxy dynamics (Sellwood and Binney 2002); moreover, mixing models were employed in evolutionary dynamics to study population genetics to search clues for mutation, selection and genetic drift (Korolev, Avlund et al. 2010).

Viscous fingering, arising in mixing process with generating dendritic or finger-like patterns, refers to the onset and evolution of instability of the contact front between two immiscible or miscible fluids when one invades the other in porous media (Homsy 1987). It has been being the most important and fascinating phenomenon occurring in mixing flows that confronts so many experimentalists, engineers, physicists, mathematicians and even computer scientists. The instability (also known as Saffman-Taylor instability) caused by the viscosity difference results in many different phase patterns or even turbulence due to its unpredictable dynamical behaviors (Drazin and Reid 2004). Fingering flows (in case of generating finger-like patterns), can be easily reproduced and approximated in the Hele-Shaw cell----a pair of transparent glass plates separated by a thin gap of an infinitesimal constant distance. For example, we may see that air invades water or water invades oil. In either case, the less viscous fluid could create finger-like patterns in the more viscous fluid. Admit that, in the real world, the mechanisms of fingering flows could be quite complicate conspired by multiple factors, such as Kelvin-Helmholtz instability due to velocity difference or Rayleigh-Taylor instability due to density difference.

Despite the pervasive existence of viscous fingering in the natural world, the systematic and mathematical studies on the viscous fingering only began in the 1950s. Hill is thought to be the pioneer in studying the phenomenon of viscous fingering (Hill 1952, Marshall and Bakker 2004). He performed experiments by diving sugar liquor in the columns of granular media filling with water and interpreted the fingering phenomenon qualitatively using a simple stability analysis. Later on, Hill continued the study on gravity-stabilized viscous fingering experimentally by including the contributions of both density and viscosity. In the late 1950s, the boundary of the viscous fingering research was pushed by Saffman and Taylor (Saffman and Taylor 1958), and Chouke *et al*. (Chouke, Van Meurs et al. 1959) for they rigorously establishing the criterion of the stability. Chouke *et al.* did the linear stability analysis on fingering flows of immiscible fluids. However, when being applied to miscible fluids, the theory was problematic due to an unbound growth constant. Since then, experiments, theories, and numerical analysis have been developing in studying fingering flows (Tan and Homsy 1988, Zimmerman and Homsy 1991, Drazin and Reid 2004, Kim and Choi 2011).

Nowadays, viscous fingering plays a significant role in many fields including but not limited to chromatographic separation of mixtures of various compounds (Rousseaux, Martin et al. 2011), exploitation of petroleum (Orr and Taber 1984), or underground water contamination. An elusive topic pertaining to viscous fingering is how to describe the degree of mixing from fingering effectively. Jha *et al.* characterized the evolution of the degree of mixing between two fluids of different viscosities using a scalar variance of the concentration field (Jha, Cueto-Felgueroso et al. 2011). They proposed scalar equations characterizing the degree of mixing in fingering flows. According to the paper, viscous fingering of miscible fluids is the result of two competing effects: enhance mixing and reducing mixing. The fingering becomes evident when the flow is disordered as the "contact area" between the two fluids increases; whereas in the meantime the formation of fingering channels in the less viscous fluid confines the other fluid of higher viscosity to sweep large areas of the less viscous fluid and hence reduces the mixing. Phenomenally, their theory links the degree of mixing from viscous fingering to its morphology of the "contact area". Later on, Chui *et al.* from the same group investigated the relationship between the evolution of interface and degree of mixing experimentally (Chui, de Anna et al. 2015). Therefore, understanding the morphology of the flow patterns is crucial to interpreting the degree of mixing.

Regardless the simplicity of the problem, the great difficulty involves the prediction of the motion of the free boundary separating two fluids (Bensimon, Kadanoff et al. 1986). For viscous fingering from immiscible fluids, numerous works have been done on the evolution of the interface. A general framework describing the interface evolution of two phases was done by Brower *et al.*, who formulated local equations of motion correlating both velocity and local curvature (Brower, Kessler et al. 1984). A dynamical model for studying singularities from a different perspective was proposed to describe the secondary tip-splitting (Elezgaray 1998). Nonlinear stability analysis was also helpful on understanding evolution of the interface (Casademunt 2004). Analytical methods based on the Buckley-Leverett equation provided an easy way to estimate the evolution of the interface (Brailovsky, Babchin et al. 2006, Babchin, Brailovsky et al. 2008). Statistical methods of a variant diffusion-limited aggregation model were also employed in studying the evolution of the interfaces (Paterson 1984, Tartakovsky, Neuman et al. 2003). Recently, concepts using the lattice Boltzmann simulation of viscous fingering of two fluids in porous media have been proposed as well (Grosfils and Boon 2003, Dong, Yan et al. 2011).

While extensive work has been done on the interface evolution of viscous fingering from immiscible fluids, seldom study has been dedicated to the dynamics of the fingers from miscible fluids. The difficulty increases when it comes to dealing with miscible fluids. Conceptually, the reason may be due to the lack of a sharp interface between two miscible fluids. However, the finger-like patterns observed in experiments are pretty much the same as that of immiscible fluids. This sort of similarity casts us a question of how to characterize the topological morphology of the viscous fingering from miscible fluids? A quick remedy is to apply numerical simulations of the governing equations and to interpret the dynamics of fingers using obtained numerical results (Tan and Homsy 1988, Coutinho and Alves 1999, Marshall and Bakker 2004). As it is known, this resolution may not cover the generality of the problems. Aside from using the advection-diffusion equation, phenomenological models were also introduced in simulating the viscous fingering growth from miscible fluids. Paterson was the first applied the diffusion-limited aggregation model to the analysis of fingering flows (Paterson 1984), of which the results showed the tip splitting occurs all the times. This discrepancy is due to random noises that may be minimized by modifying the pressure term (Chen 1988); unfortunately, the diffusion-limited aggregation model is still sensitive to the lattice size and limited to very large mobility ratios. Although by introducing molecular diffusion the BGK lattice gas method was also employed to simulate the finger growth in miscible fluids, it is not capable of capturing rich dynamical phenomena such as shielding and splitting (Rakotomalala, Salin et al. 1997). Another remedy is to perform stability analysis (Tan and Homsy 1986, Ben, Demekhin et al. 2002, Rousseaux, De Wit et al. 2007, Pramanik and Mishra 2013), but it can only predict the onset and growth of finger patterns at an early stage. It is not able to predict the finger evolution and growth on the nonlinear regime. To the best of our knowledge, the study of dynamics and deformation of the fingers involving both topological and geometrical deformations of the fingers remains incomplete. Particularly, theories on the evolution of the contact front of miscible fluids at a late stage are still blank.

In recent years, the attention has been focused on investigating the evolution of fluid-fluid interfaces. Bischofberger *et al.* studied radial finger growth experimentally and formulated a relation between viscosity ratio and finger growth rate (Bischofberger, Ramachandran et al. 2014). Chui *et al.* examined the interface evolution experimentally by defining a so-called interface length (Chui, de Anna et al. 2015). Unfortunately, it is still unclear how to effectively and accurately describe the evolution of the finger-like patterns theoretically. In this paper, we thus try to investigate the mechanism through the equations governing Hele-Shaw flows, an approximation of miscible fingering flows. The advection-diffusion equation and Darcy's law are solved using finite element methods. Finger-like patterns are generated in the simulations. The effective interfacial tension related to Korteweg stresses are then calculated, of which the magnitudes enable us to discuss the necessity of the definition of effective surface tension and interface. Finally, using the geometrical analogy of the concentration profile, the fingering phenomena in the mixing flow are readily analyzed, of which the underline physical meanings are correlated to the evolution of the concentration surface, as well as its contours (effective interface).

## 2. Formulation

We may begin with a review of the advection and diffusion mechanisms, as well as presenting a logical, rather than rigorous mathematical derivation of the advection-diffusion equation for the sake of explicit physical insight.

### 2.1. Advection and diffusion

Advection refers to the transport of fluid particles or conserved properties in the flows from one place to another along the direction of fluid velocity (streamlines). Thus, the degree of advection depends on the velocity of the flows. It can be described by the following formula in 2-dimensional Cartesian coordinate

$$\mathbf{u} \cdot \nabla = u_x \frac{\partial}{\partial x} + u_y \frac{\partial}{\partial y} \tag{2.1}$$

where $\mathbf{u} = (u_x, u_y)$ is the velocity. Denote the concentration $c$ by the ratio of the mass of a substance $m$ to the total volume of the mixture $V$, that is $c(\mathbf{x}, t) = m/V$. By assuming the incompressibility of the flow, we may write the advection equation in the following form

$$\frac{\partial c}{\partial t} + \mathbf{u} \cdot \nabla c = 0 \tag{2.2}$$

Diffusion, being another transport mechanism at the meantime, spreads substance from one place to another randomly due to the variation of the concentration gradient (or Brownian motion). This behavior may be governed by the following equation (Cussler 2009)

$$\frac{\partial c}{\partial t} - D\nabla^2 c = 0 \tag{2.3}$$

where $D$ is the diffusion coefficient for the substance in the background fluid depending on the phase, temperature, and molecule size. In order to capture both the advection and diffusion effects in the flow, the process of advection and diffusion is assumed to be linearly independent so that the principle of superposition may be imposed. Define $\mathbf{j}$ by the total flux in a control volume, in which the net flux is

$$\mathbf{j} = \mathbf{j}_{advection} + \mathbf{j}_{diffusion} = \mathbf{u}c - D\nabla c \qquad (2.4)$$

Casually, the advection-diffusion equation (ADE) is readily derived by superposing both the effects (Cussler 2009)

$$\frac{\partial c}{\partial t} + \nabla \cdot (\mathbf{u}c - D\nabla c) = 0 \qquad (2.5)$$

Incorporate the incompressible condition, and then nondimensionalize the above equation

$$\frac{\partial c}{\partial t} + \mathbf{u} \cdot \nabla c - \frac{1}{\text{Pe}} \nabla^2 c = 0 \qquad (2.6)$$

where the Peclet number is defined as $\text{Pe} = UL/D$, and $c \in [0,1]$.

It is evident that the Peclet number controls the competing effects due to diffusion and advection ($\text{Pe} \to \infty$ indicates the ADE approaching the pure advection equation, and $\text{Pe} \to 0$ suggests pure diffusion equation). For instance, the characteristic velocity $U$ relates to the advection effect and the diffusion coefficient $D$ refers to the Brownian motion. Generally, advection dominates the mixing process when $\text{Pe} \gg 1$, whereas diffusion dominates the mixing when $\text{Pe} \ll 1$.

## 2.2. Viscous fingering

The phenomenon of viscous fingering due to miscible flows in a porous medium at very low speed can be studied analogically by the flows in the so-called Hele-Shaw cell (see *Figure 1*) as long as the dispersion is isotropic (Homsy 1987).

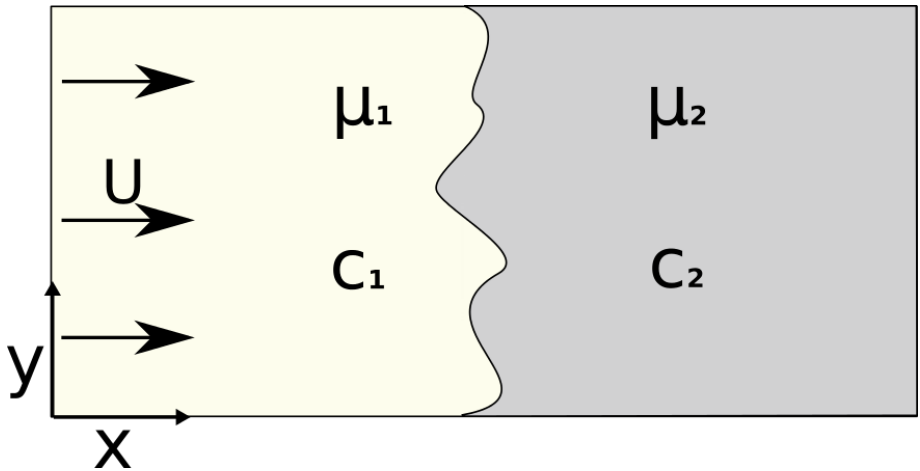

**Figure 1. Schematic representation of the flow system.**

Although complicated situations in practice may also violate the kinship, the problem of flows in Hele-Shaw cell is of great importance both mathematically and physically in its own right. Essentially, the following non-dimensional equations may be introduced

$$\nabla \cdot \mathbf{u} = 0, \quad \mathbf{u} = -\frac{1}{\mu(c)}\nabla p \tag{2.7}$$

such that the ADE (2.5) is closed. The two equations in (2.7) are the so-called Darcy's law governing flows in porous media. In the end, to characterize the viscous fingering flows, we have the following three equations

$$\frac{\partial c}{\partial t} + \mathbf{u} \cdot \nabla c - \frac{1}{\text{Pe}}\nabla^2 c = 0$$
$$\nabla \cdot \mathbf{u} = 0, \quad \mathbf{u} = -\frac{1}{\mu(c)}\nabla p \tag{2.8}$$

where the coefficients are as follows (Jha, Cueto-Felgueroso et al. 2011)

$$\mu(c) = e^{R(1-c)}, \quad R = \ln M,$$
$$M = \frac{\mu_2}{\mu_1} \quad \mu_2 = e^R, \quad \mu_1 < \mu_2 \tag{2.9}$$

The choice of viscosity-concentration relation $\mu(c)$ here follows standard literatures rather than mathematical or physical rigor (Tan and Homsy 1986, Homsy 1987, Tan and Homsy 1988, Chen and Meiburg 1998, De Wit, Bertho et al. 2005, Pramanik and Mishra 2013), admitted that the relation may be different in reality and result in different behaviors (Koval 1963, Manickam and Homsy 1993, Kim and Choi 2011). When the mobility ratio $M < 1$, the contact front of the invading fluid is stable and simple. Also, the mixing process of the mixing flow is efficient; while in realistic settings $M > 1$, the front is unstable, which results in finger-like patterns being generated in the host fluid. The fingering creates channels in which the invading fluid is confined and thereby decreases the mixing efficiency. However, it is the unstable nature gives rise to so many intricate patterns that attract so many researchers. Numerical simulation of fingering flows can be carried out on the system (2.8). Besides, the condition for 2-dimensional incompressible flows renders stream functions for the description of the velocity field; many published works employed that trick for numerical computations (Tan and Homsy 1988, De Wit, Bertho et al. 2005). In this paper, we analyze the fingering flows using the finite element method for numerical computation, of which the results are sequentially analyzed using some simple concepts of curves and surfaces in differential geometry. For the sake of convenience, Darcy's law is converted equivalently into the following Poisson equation

$$\nabla \cdot \left( \frac{1}{\mu(c)} \nabla p \right) = 0, \quad p \in \Omega \tag{2.10}$$

By expanding the above equation, we also have the following identical form

$$\nabla^2 p + R \nabla p \cdot \nabla c = 0 \tag{2.11}$$

Therefore, we finally get the following equations for the numerical analysis

$$\frac{\partial c}{\partial t} - \frac{1}{e^{R(1-c)}} \nabla p \cdot \nabla c - \frac{1}{\text{Pe}} \nabla^2 c = 0 \tag{2.12}$$
$$\nabla^2 p + R \nabla p \cdot \nabla c = 0$$

### 2.3. Pressure variation

For the preparation of mathematical analysis on the fingering flows, we may rearrange the equation (2.12) and write it in the form of

$$\nabla p \cdot \nabla c = e^{R(1-c)} \left( \frac{\partial c}{\partial t} - \frac{1}{\text{Pe}} \nabla^2 c \right)$$
$$\nabla p \cdot \frac{\nabla c}{|\nabla c|} = \frac{\mu(c)}{|\nabla c|} \left( \frac{\partial c}{\partial t} - \frac{1}{\text{Pe}} \nabla^2 c \right) \tag{2.13}$$

Equation (2.13) characterizes how the pressure changes with respect to the concentration. The left-hand side is nothing but directional derivative along the $\mathbf{n} = \nabla c / |\nabla c|$, thus define $\partial p / \partial n := \nabla p \cdot \mathbf{n}$, recalling the definition of directional derivative

$$\frac{d}{d\tau} p(\mathbf{x} + \tau \mathbf{n})|_{\tau=0} = \nabla p(\mathbf{x}) \cdot \mathbf{n} \tag{2.14}$$

we have

$$p(\mathbf{x} + \frac{\varepsilon}{2} \mathbf{n}) - p(\mathbf{x} - \frac{\varepsilon}{2} \mathbf{n}) = \int_{-\varepsilon/2}^{\varepsilon/2} \nabla p(\mathbf{x} + \mathbf{n}\tau) \cdot \mathbf{n} \, d\tau \tag{2.15}$$

Here we have applied the integration in a region perpendicular to each of the contours with a thickness of $\varepsilon$. Then, it can be written as

$$\delta p = \int_{-\varepsilon/2}^{\varepsilon/2} \nabla p(\mathbf{x} + \mathbf{n}\tau) \cdot \mathbf{n} \, d\tau \tag{2.16}$$

We can also evaluate the right-hand side of the above equation using the mean value theorem in calculus, that is

$$\begin{aligned} \delta p &= \varepsilon \nabla p(\mathbf{x} + \mathbf{n}\tau_0) \cdot \mathbf{n} \\ \delta p &= \varepsilon \nabla p(\mathbf{x}) \cdot \mathbf{n} \big|_{\tau_0 = 0} \end{aligned} \quad (2.17)$$

Finally

$$\delta p = \varepsilon \frac{\mu(c)}{|\nabla c|} \left( \frac{\partial c}{\partial t} - \frac{1}{\text{Pe}} \nabla^2 c \right) \quad (2.18)$$

which characterizes the pressure jump across the concentration contours all over the curve *c*. More analyses on the equation will be given in the coming sections.

## 3. Geometry of surfaces

The subject can be easily found in many external sources (Peng and Chen 2002, O'neill 2006, Toponogov 2006). For the convenience of reading, we briefly introduce the basic notations related to the surfaces in the 3-dimensional Euclidean space. By the explicit surface, we mean

$$z = f(x, y) \quad (3.1)$$

is a smooth function and can be differentiable to any order. We then can define the graph for the function as

$$\Gamma = \{(x, y, z) \mid z = f(x, y)\} \quad (3.2)$$

The normal vector to the surface is

$$\mathbf{n} = -\frac{(-f_x, -f_y, 1)}{(f_x^2 + f_y^2 + 1)^{1/2}} \quad (3.3)$$

Here we denote $f_x = \partial f / \partial x$, and hence $f_{xx} = \partial^2 f / \partial x^2$, $f_{xy} = \partial^2 f / \partial y \partial x$ e.t.c. Thus the two critical parameters associated with the surface are Gaussian curvature

$$K = \frac{f_{xx} f_{yy} - f_{xy}^2}{(f_x^2 + f_y^2 + 1)^2} \quad (3.4)$$

and mean curvature

$$H = \frac{(1+f_x^2)f_{yy} - 2f_x f_y f_{xy} + (1+f_y^2)f_{xx}}{2(f_x^2 + f_y^2 + 1)^{3/2}} \quad (3.5)$$

We shall see how these parameters can be used to characterize the mixing flow in the next section the area of the surface can be calculated using the following formula

$$S = \iint\limits_{(x,y)} \sqrt{f_x^2 + f_y^2 + 1}\, dxdy \quad (3.6)$$

As for implicit curves $\Gamma = \{(x, y) \mid f(x, y) = 0\}$ in plane, we may have its normal vector

$$\mathbf{n} = -\frac{(f_x, f_y)}{(f_x^2 + f_y^2)^{1/2}} \quad (3.7)$$

Moreover, the curvature

$$\kappa = \frac{f_x^2 f_{yy} - 2f_x f_y f_{xy} + f_y^2 f_{xx}}{(f_x^2 + f_y^2)^{3/2}} \quad (3.8)$$

## 4. Numerical simulation

The computational domain is $W \times H = 2 \times 1$ ($W$ in $x$-direction and $H$ in $y$-direction) as the schematic in figure 1. The aspect ratio is hence defined as $A = W/H = 2$. The less viscous fluid of viscosity $\mu_1$ enters from the left with a uniform velocity $U$, displacing the host fluid of viscosity $\mu_2$. The density gradient due to gravity is ignored. The medium is assumed homogeneous and has a constant permeability. The simulation length is relatively long enough so that the concentration and velocity at boundaries are not disturbed by the fingering effect. To mimic a porous medium, the simulation is carried out with using the following initial conditions

$$c(t=0) = \begin{cases} c_1 + \frac{1}{2} f(x, y), & x < \frac{W}{128} \\ 0, & else \end{cases} \quad (4.1)$$

where $f(x, y) \in [0,1]$ is a random function serving as the random disturbance of the initial concentration. The numerical simulations are well documented using other numerical techniques, such as finite difference methods (Christie, Muggeridge et al. 1993, Chen and

Meiburg 1998) and pseudo-spectral methods (Tan and Homsy 1988, De Wit, Bertho et al. 2005). We use finite element method to discretize and solve the equation system (2.8). Neumann conditions for the Poisson equation and Dirichlet conditions for the advection-diffusion equations are applied; as for the upper and bottom boundaries, the periodic boundary condition is employed.

## 5. Results

The results from numerical simulations are presented by the four snapshots in *Figure 2*, which demonstrate a wealth of finger-like patterns involving the phenomena of spreading, shielding, and coalescence. The initial disturbance was simulated using the random function (4.1) mimicking flows in porous media. At the early stage, the instability was initially triggered by the disturbance, and then finger-like patterns were developed and grown. The envelope of the fingers in the last snapshot resembles a parabolic shape. Another interesting observation in a numerical simulation with anisotropic dispersion and moderate Peclet number Pe = $10^3$ is the phenomenon of fading (Zimmerman and Homsy 1991). However, without the anisotropic dispersion being assumed in the present simulations, here we can still see from *Figure 2* that the fading phenomenon occurs when the less viscous fluid chooses to flow preferentially through the adjacent finger channels. A similar result was also captured by other simulations with the isotropic dispersion (Jha, Cueto-Felgueroso et al. 2011). Similarly, the phenomenon of coalescence (finger merging) seen in the numerical simulations with the anisotropic dispersion, is also observable in *Figure 2* (see also the supplemental movie). Since neither of the present nor Jha's simulations considered the impacts from the anisotropic dispersion, we think that the anisotropy of the dispersion is not the trigger to the fading/coalescence phenomenon. We also found another discrepancy in numerical simulations: the parameter set in the present simulations are similar to that of in Jha *et al.* (particularly, the Peclet number is the same), but the tip splitting does not occur in the present simulation; meanwhile, the simulations done by Zimmerman and Homsy did not always produce the tip splitting either, even though they used a similar set of parameters suggested by the results from linear stability analysis (Zimmerman and Homsy 1991). The onset of tip splitting (the secondary instability) that are foreseeable with the linear stability analysis may not be directly affected by the Peclet number either (Tan and Homsy 1988). It is known that, for immiscible fluids, the surface tension inhibits the onset of the tip splitting (Chuoke, Van Meurs et al. 1959). The fingering flow of miscible fluids with a large Peclet number could also give rise to the phenomenon of transit surface tension. Thus, the large Peclet number in the simulation may decrease the possibility of the onset of the secondary instability and hence supports the current results. But Jha's observation on tip-splitting remains unexplainable. The above discrepancies suggest that the Peclet number may not be the dominant factor in determining the nonlinear phenomena, e.g., coalescence and tip splitting, and the instability is subtle.

Multiple factors may conspire to the same phenomenon of the nonlinear fingering, instead of doing separately (Malhotra, Sharma et al. 2015).

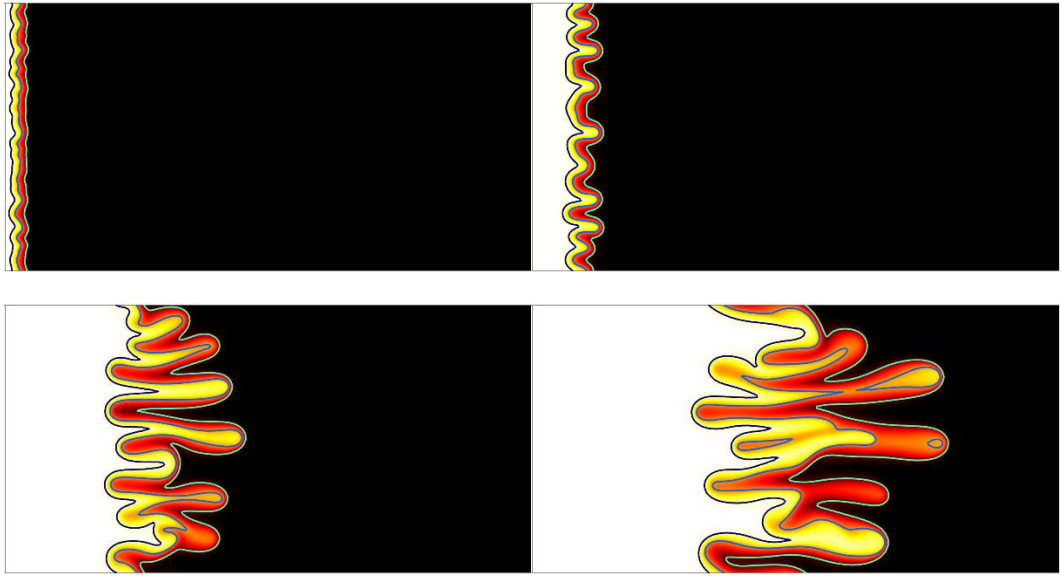

Figure 2. The concentration contours of the flow pattern for Pe = $10^4$, R = 2, A=2, and t = 1, 3, 8, 12, respectively.

To see how the concentration field varies dynamically, we plot the evolution of averaged concentrations in *Figure 3* as well. The two set of data were obtained by averaging the concentration over the entire computational domain (blue solid curve) and the mixing zone (area enclosed by the contours of 0.1 and 0.9, red dashed curve). The increasing tendency of both curves is similar, except for a larger slope of the concentration averaging over the domain due to more portions of the high concentration fluid included. The purpose of displaying this result is to justify that there may be no critical difference in investigating the mixing flow between the two different averaging methods. Therefore, we are able to study the characteristics over the whole domain for the matter of convenience. This justification will be enforced further in the coming sections.

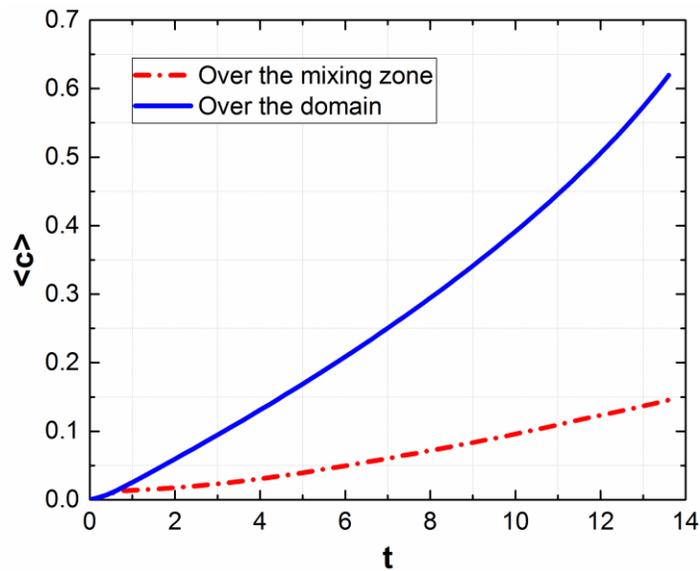

Figure 3. Average concentration vs. time.

### 5.1. Qualitative observation

In order to understand the morphology or the dynamic evolution of the finger-like patterns, we qualitatively analyze the concentration profile with respective to the pressure, velocity, and concentration gradient.

#### 5.1.1. Pressure field

A pressure jump across the interface of two immiscible fluids is known to be reasonable due to the existence of an equilibrium interface. However, it is not the case for miscible fluids because no equilibrium interface exists. This fact can also be seen from the coupled equation system, in which the pressure field induced by the concentration variation is continuous and smooth. The pressure jump thus seems hardly exist. Nevertheless, the finger-like patterns are remarkably similar to that of immiscible fluids. In *Figure 4*, we plot the concentration contours at $c = 0.1$ (blue curve) and $c = 0.9$ (red curve) with adding the pressure field in the form of parallel lines. *Figure 4* shows how the three quantities -- pressure gradient, pressure, and concentration --of the coupled system interact with each other.

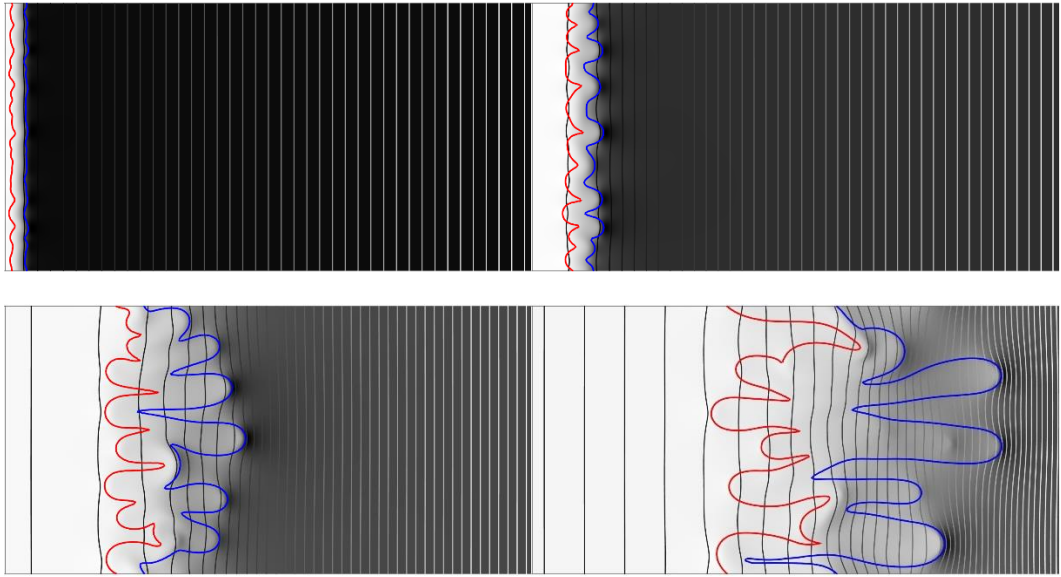

**Figure 4. The pressure gradient (intensity of the white shades), pressure contours (parallel lines), and concentration contours (blue and red curves) of the fingering flow for Pe = 104, R = 2, A=2, and t = 1, 3, 8, 12, respectively.**

Initially, the pressure contours are in the form of parallel lines in the first graph of *Figure 4*, indicating that the initial disturbance only affects the concentration field locally. Then, with time evolving, the distortion of the concentration contour is being amplified (growth of the mixing zone), and hence gives rise to the deformation of the pressure contours. The pressure thus is not a constant transverse to the direction of the longitude growth of the fingers in the mixing zone. From this perspective, we may feel the temporal fingering instability is due to the unsteady pressure profile. Intuitively, the fingers push forward the pressure contours so that those in front of the tips curve more. The pressure contours tangent to the fingertips and bend in the same direction as the corresponding fingers, whereas the situation reverses at the finger roots. Those pressure contours in the middle of the mixing zone are more irregular than others, which closely links to the degree of fluid mixing. The curvatures of pressure contours in this zone may also be related to the nonlinear fingering phenomena. A fact is the pressure disturbance exists only in the mixing zone of the flow. After the mixing zone propagating forward to the right, the pressure field of the invading fluid at the left behind restores to its regularity. However, the density of the contour lines decreases due to the low viscosity of the invading fluid. The gray scale color resembles the gradient of the pressure field.

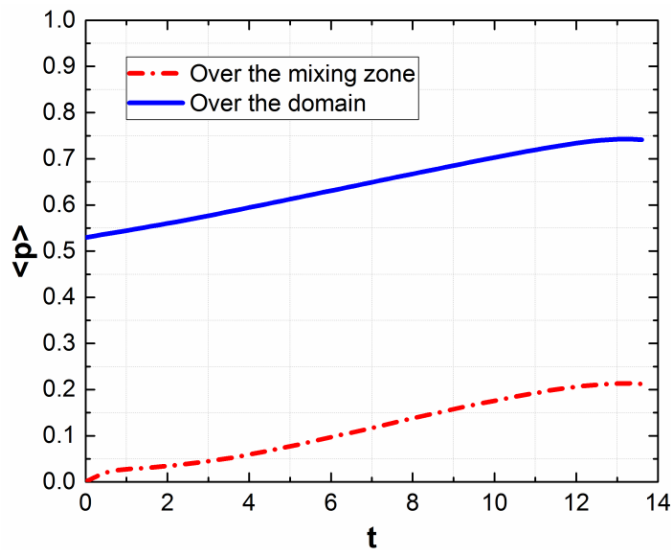

**Figure 5. Average pressure vs. time.**

*Figure 5* demonstrates how the averaged pressure evolves with respect to time. As we have already seen in *Figure 3*, the tendency is pretty much the same. It again suggests that we can just investigate the mean values over the entire domain for simplicity.

*5.1.2. Velocity field*

To better track the movements of the mass and characterize the finger morphology, we draw the streamlines in the flow field in *Figure 6*. As we can expect, the streamlines deform only in the mixing zone because outside the mixing zone, the flow is steady, and the streamlines do not change with respective to time, as opposed to the unsteady flow in the mixing zone due to the concentration variation. Yortsos *et al.* defined a transverse equilibrium regime, in which the pressure gradient is independent of *y*-coordinate and the flow is parallel (Yortsos and Salin 2006). However, this regime has not been developed in our simulations since both pressure contours and streamlines are always curved in the mixing zone. The curvature of a streamline relates to the pressure gradient acting perpendicular to the streamline. The direction of the curvature is along the same direction as the pressure gradient. A steeper pressure gradient results in more curved streamlines, which are observable when a finger is spreading (growing in *y*-direction). The density of the streamlines is proportional to the velocity. As we can see that inside fingers, the streamlines get higher densities, suggesting that the velocity of the fluid inside the fingers is higher than that of outside the fingers. Phenomenally, therefore, we can conclude that the curvatures of streamlines determine the transverse growth of a finger while the density of the streamlines inside it controls the longitude propagation. Since the velocity inside a finger is higher that of outside the finger, the pressure inside is lower. It thus increases the possibility of sucking nearby fingers, resulting in coalescence.

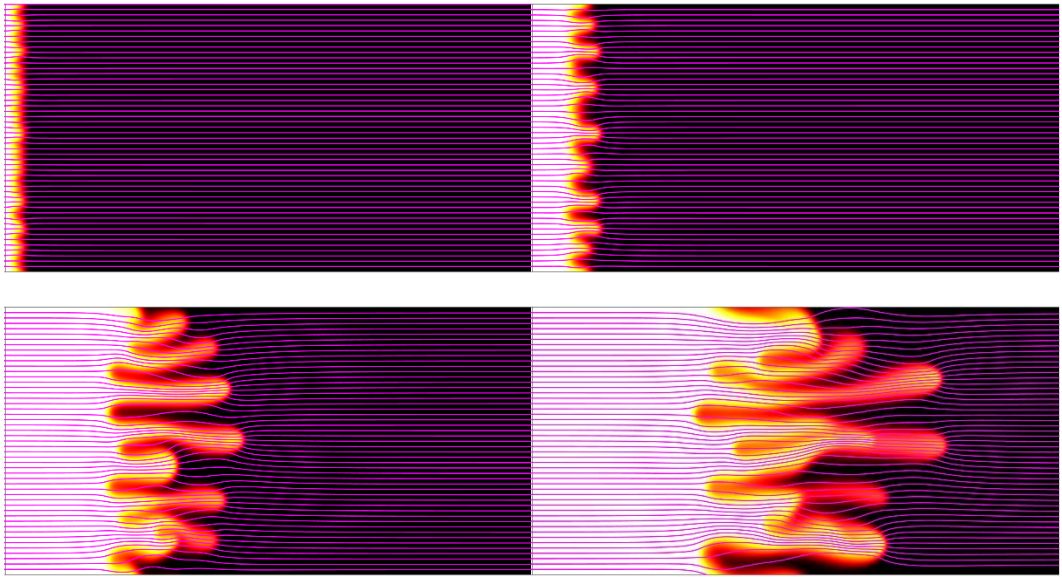

**Figure 6. The concentration patterns and streamlines of the fingering flow for Pe = 104, R = 2, A=2, and t = 1, 3, 8, 12, respectively.**

The last graph in figure 6 demonstrates the phenomenon of coalescence, happening at which the streamlines becomes much denser.

### 5.1.3. Growth of Mixing zone

Experiments and theories focusing on the growth of the mixing zone of miscible fluids are well documented. Although the theory of growing mixing zone is incomplete (Menon and Otto 2005, Yortsos and Salin 2006, Booth 2010), numerical investigations are easy to find. By defining the entropy of the mixing zone, the growth can also be calculated by the entropy solutions (Menon and Otto 2005, Yortsos and Salin 2006). Here we follow the convention to discuss the growth of the mixing zone of miscible fluids, and the mixing zone is denoted by $\Omega(t) = \{(x, y) | 0.1 \leq c(x, y, t) \leq 0.9\}$. We plot the concentration contours of 0.1(blue) and 0.9 (red) in *Figure 7*.

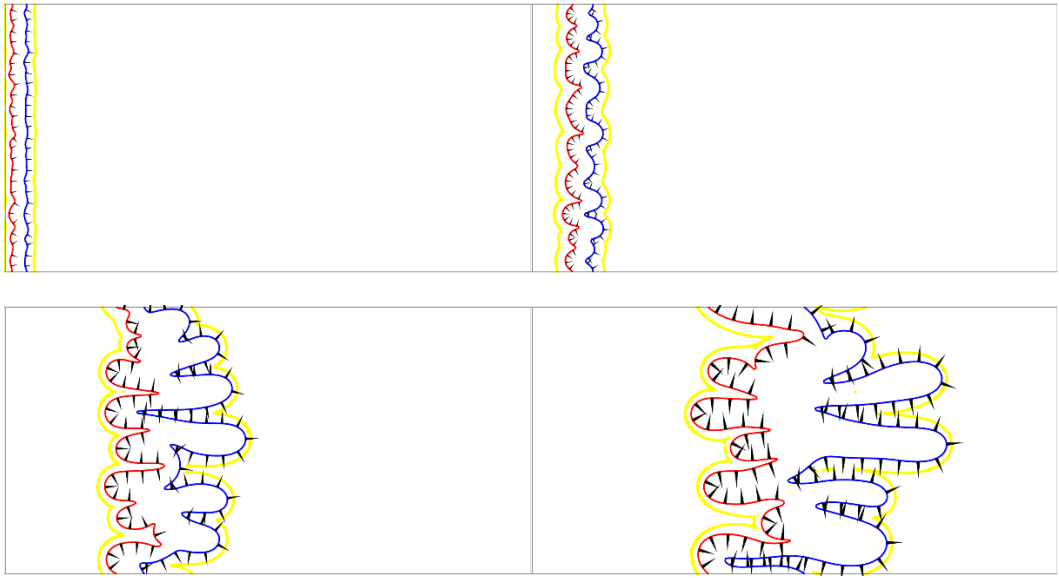

**Figure 7. The contours of concentration gradient at 0.1 (yellow) and concentration contours at c = 0.1 (blue curve) and c = 0.9 (red curve) of the fingering flow for Pe = 104, R = 2, A=2, and t = 1, 3, 8, 12, respectively. The cones are normal vectors of the concentration contours.**

Data from both experiments and mathematical analyses showed that the rate of the growth of the mixing zone is proportional to the time *t* (Koval 1963, Malhotra, Sharma et al. 2015). Additionally, for a diffuse interface the growth of the interfacial thickness, as well as the width of a finger are proportional to the square root of the time (Wooding 1969, Menon and Otto 2005). The degree of mixing was also defined as a function of areas of the mixing zone. From *Figure 7* we can see the fingers can affect the mixing process. It is because the formation of fingers decreases the degree of mixing by restricting the invading fluids into the finger channels and thereby reducing their contact areas. The actual mixing zone is thus the area between the invading (blue) and invaded (red) fingers. From this perspective, a method to enhance the mixing is to enlarge the distance of the two contours or the area of the zone. An alternative definition of the mixing zone can also be defined by imposing $|\nabla c|=0.1$. It is enclosed by yellow curves in *Figure 7*. The evolution of the mixing zone for both the definitions is presented in *Figure 8*.

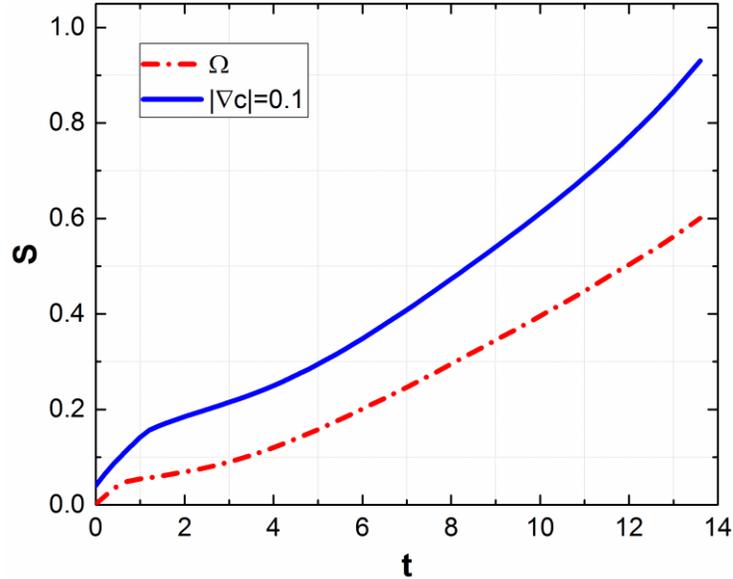

**Figure 8. The growth of the mixing zone defined by two different methods vs. time.**

### 5.1.4. Concentration gradient

As already mentioned before, the concentration gradient is the key to driving the mixing flow. The averaged concentration or pressure may be not capable of characterizing the dynamical details in the mixing zone since they average out to many details. Therefore, we visualize the impact of the concentration gradient on the evolution of the contours by drawing the concentration gradient onto the contour graphs as shown in *Figure 9*. The first snapshot in *Figure 9* shows that the gradients growing inside the red and blue contours are due to the initial disturbance. The gradients then move forward to the right with time evolving, which is the result of the concentration contrast against the host fluid. *Figure 3* and *Figure 5* shows that the averaged concentration and pressure are different in quantity; in contrast, the gradients of concentration are very similar to each other as shown in *Figure 10* regardless of averaging over the mixing zone or the entire domain. It is mainly because the gradient is zero outside the mixing zone. An interesting phenomenon as discussed before is that the evolution of the averaged concentration gradient mimics the evolution of the area of the mixing zone (see blue lines in *Figure 8* and lines in *Figure 10*). Therefore, a closed relationship between the dynamics of concentration and the area of the mixing zone could be built, e.g. $S(t) \propto \langle |\nabla c(t)| \rangle$. We can observe in the fourth graph of *Figure 9* that the signatures of coalescence, shielding, and fading are recovered by the concentration gradients. Therefore, the gradient is the key to understanding the process of viscous fingering.

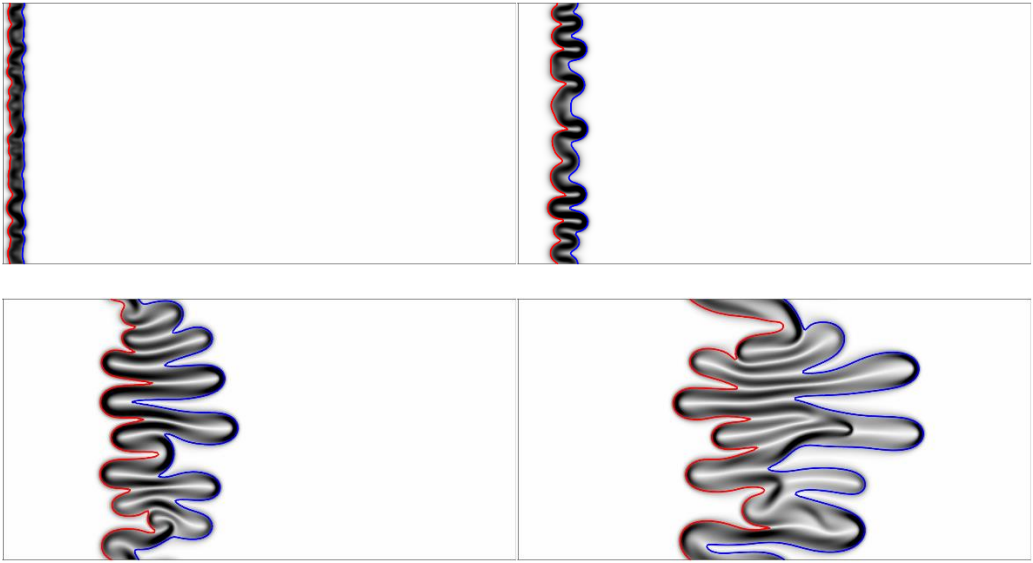

**Figure 9. The concentration contours (blue and red curves) and gradients (scale of gray color) of the fingering flow for Pe = 104, R = 2, A=2, and t = 1, 3, 8, 12, respectively.**

The abovementioned qualitative observations suggest that interesting quantities (averaged concentration, pressure, and gradients of concentration) are independent of the definitions of different mixing zones. It enables us to calculate those quantities in the whole domain, and thereby dramatically simplifies theoretical analysis. It also allows us to build mathematical models on the morphology the fingering patterns. Thus, the analysis in the next sections is carried on in the whole domain.

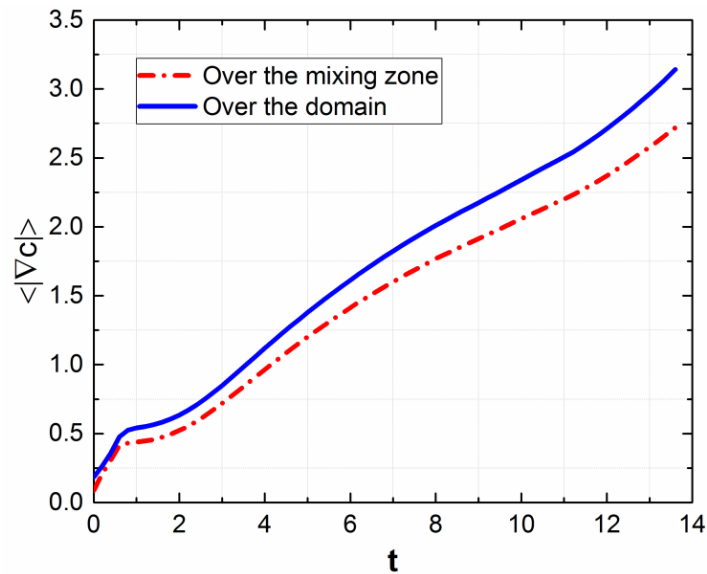

**Figure 10. Gradients of the concentration averaged over two different domains vs. time.**

### 5.2. Effective interface

#### 5.2.1. Effective interfacial tension

Traditionally, surface tension exists at the interface between two immiscible fluids, whereas it is impossible for miscible fluids due to the formation of a homogeneous mixture and hence the lack of an equilibrium interface. However, Korteweg postulated there could exist transient interfacial tension for miscible fluids (Korteweg 1901). It is due to the gradients of composition that lead to anisotropic forces which mimic the effects of interfacial tension (Joseph and Renardy 1993). Since the postulation, the debate regarding on the existence of effective interfacial tension has never ceased. Recent experiments have reported the existence of transient interfacial phenomena in miscible fluids analogous to that of observed in immiscible fluids (Pojman, Bessonov et al. 2007, Zoltowski, Chekanov et al. 2007, Truzzolillo, Mora et al. 2014, Truzzolillo and Cipelletti 2016, Truzzolillo, Mora et al. 2016). It is referred as the effective interfacial/surface tension and may be linked to the flows with large Peclet numbers (flows of the high concentration gradient and low diffusion rate). The stress due to the effective surface tension is termed Korteweg stress (Korteweg 1901, Joseph and Renardy 1993, Pramanik and Mishra 2013). Numerical experiments reported that the finger-like patterns in miscible fluids of great Korteweg stress are very much similar to that of observed in immiscible fluids, suggesting the effective interfacial tension is equivalent to its immiscible counterpart (Chen, Chen et al. 2006).

A legitimate concern associated with the above simulations is the presumption of the effective surface tension. Thus the surface tension mimicking phenomenon may be not due to the nature of the fingering flows. Nevertheless, we have already known that the composition heterogeneities at the concentration front generate pressure anisotropies and hence stress at the mixing zone between the contacted fluids. Note also that, without considering Korteweg stresses, numerical simulation using BGK methods suggested an interface between miscible fluids similar to that of immiscible counterpart may be defined for the case of high Peclet numbers (Rakotomalala, Salin et al. 1997). By assuming a small gradient of concentration of local equilibrium, the definition of the effective interfacial tension may be obtained by minimizing the free energy of the phase (locally)

$$\gamma_e = k \int_{-h/2}^{h/2} (\nabla c)^2 dx$$

Where $k$ is Korteweg constant, $h$ is the length of the mixing zone. The above effective interfacial tension was defined by assuming a planar mixing layer, which may not be proper in studying fingering flows since the instability of fingering flows results in irregular shapes of the front and back (finger-like patterns have variable curvatures). Furthermore, it is difficult to deduce the Korteweg constant *k*. Therefore, we need to find an alternate definition. It is well known that, for immiscible flows, the surface tension suppresses the instability. Numerical results also showed that the tip splitting wound not occur if the Korteweg stress added into the equation system (Pramanik and Mishra 2013). In the present simulation, the tip splitting is absence, and hence we may expect a reasonable magnitude of the Korteweg stress to exist. In order to check whether it is true or not, we need to calculate the effective surface tension. Although the present simulation does not include the Korteweg stress, we can still evaluate the effective interfacial tension by balancing the stresses around the fingers. An easy way to estimate the effective surface tension is to employ the analogy of its original definition in the following form

$$\Delta p = 2H\gamma = \kappa\gamma$$

The above expression will give us more explicit meanings if it is written in terms of equations in the previous discussion (see section 2.3)

$$\delta p = \nabla \cdot (\frac{\nabla c}{|\nabla c|}) \frac{\varepsilon \frac{\mu(c)}{|\nabla c|}\left(\frac{\partial c}{\partial t} - \frac{1}{\text{Pe}}\nabla^2 c\right)}{\nabla \cdot (\frac{\nabla c}{|\nabla c|})}$$

Since $\mathbf{n} = \nabla c / |\nabla c|$, we immediately have

$$\gamma_e := \frac{\varepsilon \frac{\mu(c)}{|\nabla c|}\left(\frac{\partial c}{\partial t} - \frac{1}{\text{Pe}}\nabla^2 c\right)}{\nabla \cdot (\frac{\nabla c}{|\nabla c|})}$$

Theoretically, $\gamma_e$ should be constant for a sharp surface. The above equation may be approximated into the former definition. Therefore, to evaluate the magnitude of the effective surface tension numerically, we can write it in the form of

$$\gamma_e = -\frac{\varepsilon \nabla^2 p}{R|\nabla c|\nabla \cdot (\frac{\nabla c}{|\nabla c|})}$$

A reasonable simplification is to assume a homogeneous state in which $|\nabla c| = \text{const}$, yielding the following simplified expression

$$\gamma_e := -\frac{\varepsilon}{R}\frac{\nabla^2 p}{\nabla^2 c}$$

which correlates to the diffusion of both concentration and pressure. The interfacial tension defined in this way may not be constant as long as the diffusion of the pressure and concentration is off equilibrium state. Thus, the effective interfacial tension may only exist locally and transiently. Adhering to the terminology of "effective", we may average it over the entire domain. Integrating the above relation, we have

$$\langle \gamma_e \rangle := -\frac{\varepsilon}{R} \frac{\iint_{W \times H} \nabla^2 p \, dxdy}{\iint_{W \times H} \nabla^2 c \, dxdy}$$

The surface tension, serving as a physical constant, plays an essential role in immiscible fluids; however, the analogy is not prominent when it comes to the miscible fluids due to the variable nature (see *Figure 11*). On the other hand, the stress due to the effective interfacial tension can be determined solely by the concentration gradient of the mixing fluids, and thus it is not an intrinsic parameter. For simple mixing fluids, the definition may have sort of significance due to the regularity of the effective interface (Pojman, Bessonov et al. 2007, Truzzolillo and Cipelletti 2016), e.g. the constant Gaussian curvatures or concentration gradients at mixing front; whereas the physical insight needs further investigation when applying it to fingering flows in which chaotic mixings reduce the homogeneity of the concentration gradients and hence results in variable Gaussian curvatures of the front.

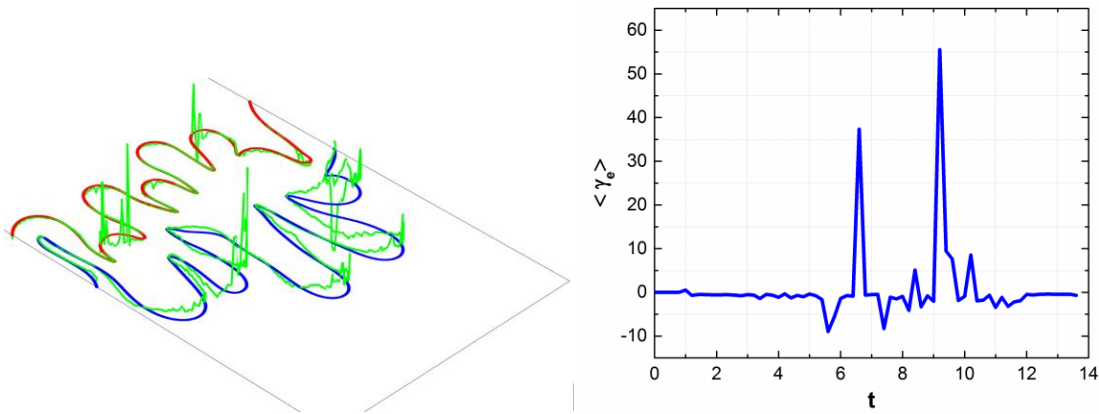

**Figure 11. The demonstration of effective surface tension on the finger ridges at t = 12, and the average effective surface tension vs. time, respectively.**

### 5.2.2. *Effective interface*

The effective interfacial tension has been defined in an *ad hoc* way, of which the values are not constant at the concentration contours. Therefore, it is not a trivial task to determine an effective interface corresponding to the interfacial tension. As we have seen in the above figures, the mixing zone, forming finger-like patterns, does resemble a distinct interface. Chui *et al.* defined the effective interface at where the maximum of the concentration gradient locates. Note that this definition may result in complex geometries due to the complexity of the concentration patterns inside the mixing zone (see ***Figure 9***). In the present simulations, although the effective surface tension is not significant at which finger-like patterns are appearing, we do see the finger-like patterns similar to that of the immiscible counterpart. Therefore, phenomenally, we may still define the location of the effective interface as the skeleton of the fingers at the concentration of 0.1 and 0.9, of which the values were used popularly among other authors (Zimmerman and Homsy 1991, Chen and Meiburg 1998). Admitted that the choice of the values may be different (Tan and Homsy 1988, Rakotomalala, Salin et al. 1997, Pramanik and Mishra 2013, Chui, de Anna et al. 2015), but it should not affect the overall characteristics (See also ***Figure 7***). Since the selection of effective interface does not suffer any topological change (such as self-intersection and splitting), it also provides a good representation of the finger-like patterns. We choose to follow such definition as the matter of convenience in geometrical treatment.

On the other hand, the concentration $c(x,y,t)$ can be seen as level surfaces at different times $c(x,y)|_t$ geometrically. Thus the ADE (2.6) becomes the level set equation of surface evolution (see ***Figure 12*** for demonstrations).

From this view, the ADE is a variant of the equation of mean curvature flows (Colding, Minicozzi et al. 2015). The concentration contours are defined as the following level sets (blue and red curves in ***Figure 12***)

$$s(t) = \{(x,y) \,|\, c(x,y,t) = 0.1\}$$
$$s(t) = \{(x,y) \,|\, c(x,y,t) = 0.9\}$$

Hence, it is not difficult to see that the gradient of the concentration is precisely the normal vectors of the implicit level curves. To this end, we have just built a relationship between physics and geometry. This shifting enables us to get more insightful interpretations of the mixing phenomenon.

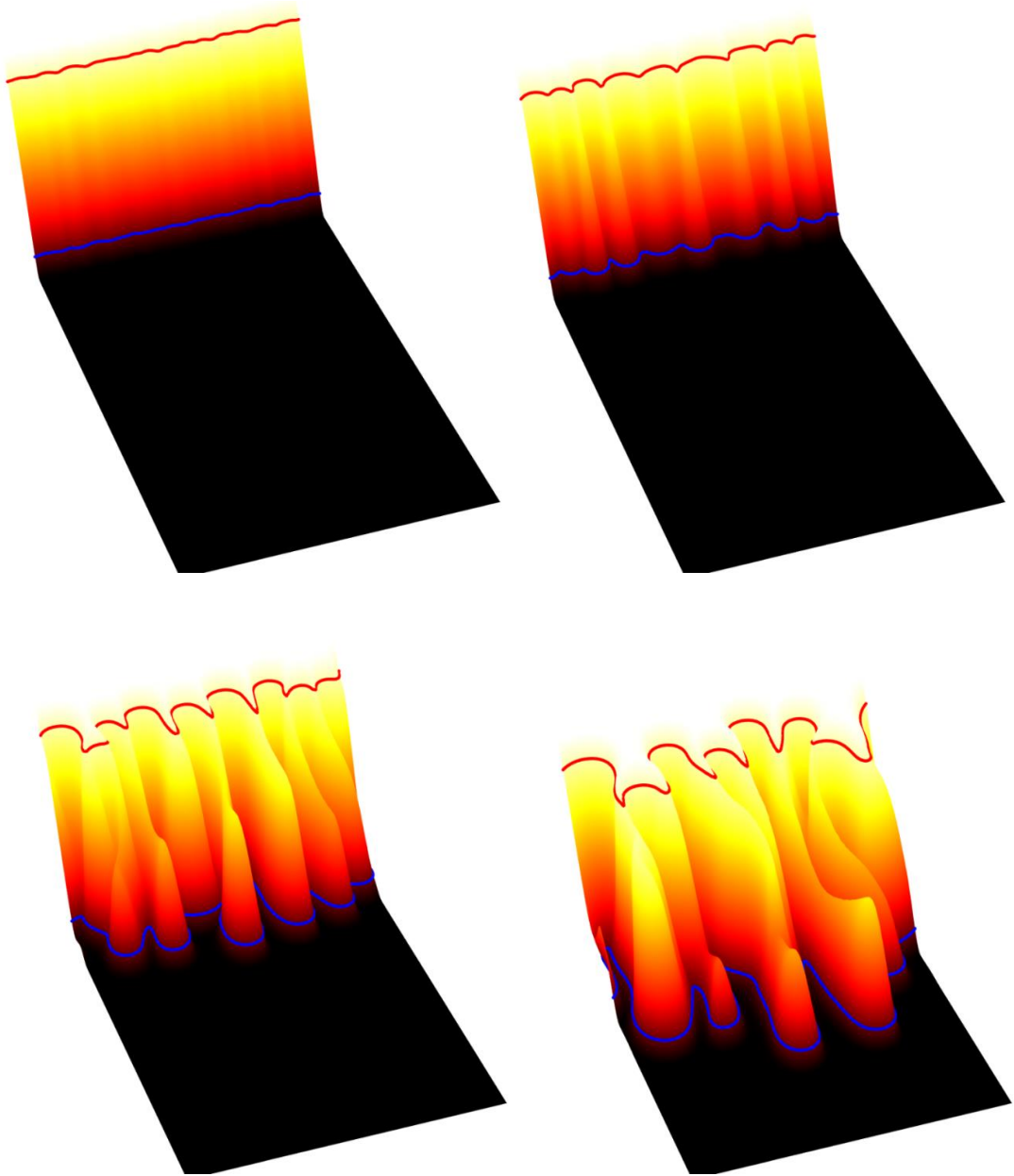

**Figure 12. The concentrations of the fingering flow are plotted as 3-dimensional level surfaces. Here again Pe = 104, R = 2, A=2, and t = 1, 3, 8, 12, respectively.**

### 5.3. Deformation of the effective interface

The discussions in section 5.1 suggest that integration over the mixing zone can be expanded into the whole domain; therefore, By seeing the concentration profile as a surface in 3-dimensional Euclidean space $\mathbf{R}^3$, we are able to investigate the deformation of the effective interface in the geometrical setting.

### 5.3.1. Length of the effective interface

The length of the effective interface and the degree of mixing may have an inverse proportional relation. The more the finger-like pattern forms, the larger the interfacial length is resultant because finger channels restrict the mixing of the fluids. The length of the effective interface is calculated accurately by extracting data from the numerical results. Alternately, recall the so-called coarea formula (Osserman 1978)

$$\iint_{W \times H} |\nabla c| \, dxdy = \int_0^\infty l(a) \, da$$

where $l(a)$ is the length of the level curves of c, namely $c(a) = \{(x, y) \,|\, c(x, y)|_t = a\}$. As the matter of convenience, we may define the length of the effective interface in the following form with possible violation of mathematical rigorous (Chen and Meiburg 1998)

$$L(t) = \iint_{W \times H} |\nabla c| \, dxdy$$

Actually, the values of $L(t)$ represent the sum (or integration) of all the level curves, instead of a single level curve at $c = 0.1$ or $c = 0.1$. On the other hand, from *Figure 13* (Left) we can easily see that the curve is very much similar to the area and pressure gradient in *Figure 8* and *Figure 10*. The geometrical result of the above length formula is pertinent to the area of the mixing zone in $\mathbf{R}^3$. Thus, the interfacial length can be investigated using the aforementioned parameters that give us more explicit geometrical and physical meanings. However, the $L(t)$ is not exactly proportional to $t^{1/2}$, which indicates a more nonlinear phenomenon existing in the mixing flows. Meanwhile, we may directly define the area of the mixing zone enclosed by the concentration contours in $\mathbf{R}^2$ physically, as well as the explicit 2-dimensional surface described by $c(x, y, t)$ in $\mathbf{R}^3$. Recall the normal vector and mean curvature of the explicit surface represented by the concentration c.

$$\mathbf{n} = -\frac{(-c_x, -c_y, 1)}{\sqrt{c_x^2 + c_y^2 + 1}}, \qquad K = \frac{c_{xx} c_{yy} - c_{xy}^2}{(c_x^2 + c_y^2 + 1)^2}$$

$$H = \frac{(1+c_x^2) c_{yy} - 2 c_x c_y c_{xy} + (1+c_y^2) c_{xx}}{2(c_x^2 + c_y^2 + 1)^{3/2}}$$

of which the values are calculated using the results from numerical simulations.

The area of the mixing zone for 2-dimenasioanl and 3-dimensional cases are simply

$$S(t) = \begin{cases} \iint_\Omega dxdy, & 2D \\ \iint_{W\times H} \sqrt{c_x^2 + c_y^2 + 1}\, dxdy, & 3D \end{cases}$$

However, neither the growth of the mixing zone of 2-dimensional nor 3-dimensional is strictly proportional to time, as we may see in **Figure 13** (Right). It suggests that the simple averaging technique for the concentration may rule out certain fine details, which could be important in predicting nonlinear fingering phenomena. Furthermore, by comparing 3D curve in **Figure 13** (Right) and $L(t)$ in **Figure 13** (Left), we are able to see that the interfacial length is a good approximation of the 3-dimensional concentration profile.

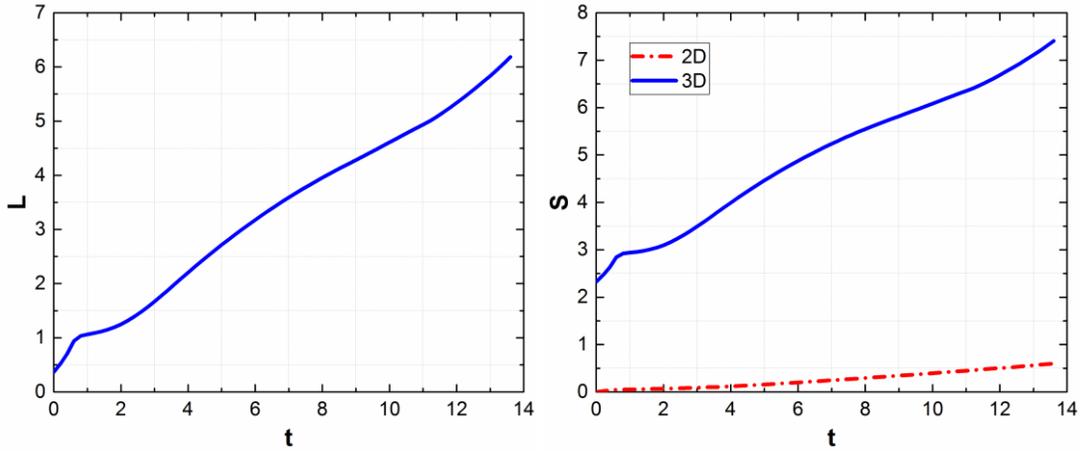

**Figure 13. The length of effective interface (Left) and Evolution of the mixing zone (Right) vs. time.**

### 5.3.2. The total absolute curvature of the effective interface

To further understand the geometrical and topological structures of the concentration surface, we calculate the total absolute curvature (also called phase distance) using (Wolfgang 2006)

$$K = \iint_S \mathrm{K}\, dS = \iint_{W\times H} \mathrm{K}\sqrt{c_x^2 + c_y^2 + 1}\, dxdy$$

The Euler characteristic can be calculated immediately in terms of Gauss-Bonnet theorem

$$\iint_{W \times H} K\sqrt{c_x^2 + c_y^2 + 1}\, dxdy = 2\pi(\chi - 1)$$

$$\chi = \frac{1}{2\pi} \iint_{W \times H} K\sqrt{c_x^2 + c_y^2 + 1}\, dxdy + 1$$

where the Euler characteristic $\chi = 1$ for a simply connected compact surface (homeomorphic to the topological disk), given that the geodesic curvature vanishes, and the rotation of the boundary at a vertex is $2\pi$. Thus, the total Gaussian curvature of the surface is zero, and a non-constant Euler characteristic introduces a nice way to investigating the topology of the concentration surface. As we can see it from *Figure 14*, in the beginning, the Euler characteristic is negative values, which corresponds to the initial perturbation of the concentration field generating many holes on the concentration surface in the region of $x < L/128$. Then, with the developing of the fingering flow, the Euler characteristic peaks at $t = 0.2$, which indicates the concentration is spreading, disjoint surfaces are appearing. After that, the curve decreases to $\chi = 1$, suggesting the concentration surface recovers its regularity and forms almost a connected saddle-like shape geometrically. The minimal value locates at $t = 11.2$. This observation may be linked to the set-on of the finger-like patterns (First instability). The negative values of $\chi$ may associate with a surface of the hyperbolic metric, or the disconnection of the concentration surface (self-intersection). Physically, when the negative values appear, holes in the flows also form. Therefore, channels in the flows are also formed, and the mixing is restricted.

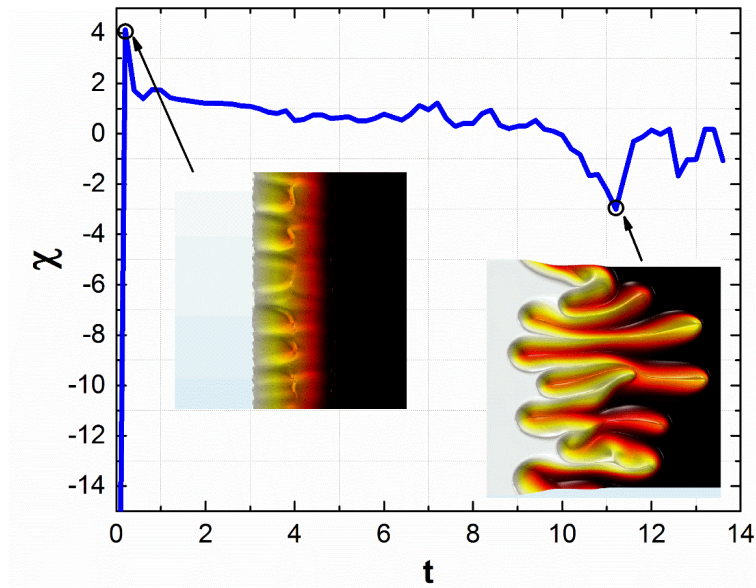

**Figure 14. Euler characteristic of the concentration surface vs. time.**

### 5.4. Degree of mixing and interface deformation

Previous studies have shown that the interface between two fluids is closely related to the deformed interface. To explore the relation between them, we at first investigate the degree of mixing defined by $\sigma^2 = \langle c^2 \rangle - \langle c \rangle^2$ (Jha, Cueto-Felgueroso et al. 2011), here $\langle \cdot \rangle$ denotes spatial averaging over the domain. Another interesting parameter thus can also be defined in terms of the isoperimetric inequality $4\pi S(t) \leq L^2(t)$ (Osserman 1978), or the Sobolev inequality as a more general case in a 2-dimensional domain (Aubin and Li 1999), thereby we introduce the Sobolev quotient as follows

$$\Sigma(t) := 4\pi \frac{\iint_\Omega c^2 dxdy}{(\iint_\Omega |\nabla c| dxdy)^2}$$

which is closely related to the degree of mixing. The perimeter inequality states that, geometrically, given a closed curve of length $L$, the largest area it can be enclosed is $L^2$. Therefore, a well-mixed medium may have smaller differences; thereby the value will top at 1. In this sense, the above formula is well defined. Furthermore, for non-radial viscous fingering, the values are always below 1, whereas it could reach 1 for radial fingering flows. The formula gives the upper bound for the degree of mixing and suggests a possible way to increase fluid mixing, e.g., using radial flows instead of rectangle flows. The dynamical behaviors in the mixing flows can also be dug out from the Sobolev ratio. For example, initially in *Figure 15* (Right), the area (the numerator in the above Sobolev quotient) grows faster than the perimeter (the denominator in the above Sobolev quotient) and then saturates at $t = 2.2$. After the peak formed, the fingering flow favors the growth of the perimeter, and thus the curve decreases rapidly. At $t = 6$, the growth rate of the perimeter starts going down. When the finger-like patterns being formed and developed, the $\Sigma(t)$ decreases substantially, suggesting that the appearing of finger-like patterns does decrease the degree of mixing. On the other hand, we can readily conclude that the degree of mixing cannot exceed 1. The degree of mixing reaches the peak at $t = 2.2$. In this perspective, we see the Sobolev ratio provides an excellent parameter to characterize both the mixing process and fingering flow.

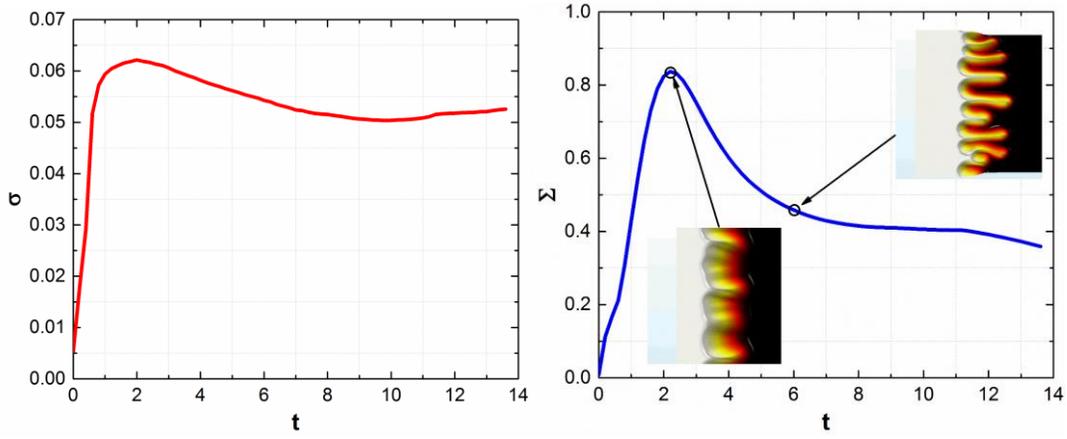

**Figure 15. The degree of mixing σ (Left) and Σ (Right) vs. time.**

## 6. Conclusion

We started the investigation with reviewing the development of mixing flows. We discussed the numerical results from the finite element simulation. A geometrical analogy from surfaces in differential geometry was brought into the current study on the morphology of the flow patterns generated by viscous fingering from miscible fluids. We derived a formula for the evaluation of the effective surface tension, of which the results were calculated at the Peclet number of Pe=$10^4$. The variable nature of the surface tension makes its physical insight elusive. Thus, the significance of the defined Korteweg stress, related to the concentration gradient, needs further consideration. Notwithstanding the above uncertainty, the effective interface does exist and resembles its immiscible counterpart. We were still able to investigate the morphology of the finger-like patterns by defining an effective interface enclosed by the concentration contours of 0.1 and 0.9. By the comparison of the averaged parameters over the entire domain and mixing zone (region enclosed by the effective interface), we found that characteristics of the fingering flow in the mixing zone could be expanded into the entire domain without affecting the results. This observation brought us convenience for quantitative analysis. When seeing the concentration profile as a 2-dimensional surface, we were able to interpret the mixing flow using geometrical parameters, avoiding the difficulties in defining the effective interface. The Euler characteristic introduced in the present study successfully explained the relationship between the morphology and the mixing phenomenon of the fingering flow. The Sobolev ratio was introduced as a mathematical means for the characterization of the fingering flow, as well as the degree of mixing.

# Supplementary

Supplementary movie is available at online version of the paper.